\pdfoutput=1
\documentclass[aps,pre,reprint,groupedaddress,twocolumn]{revtex4-1}
\usepackage{acronym}
\usepackage{color}
\usepackage{amsmath,amssymb}
\usepackage{graphicx}
\usepackage{grffile} 
\usepackage{footmisc}
\usepackage{hyperref}
\usepackage{bbold}
\usepackage{lipsum}

\newcommand{\K}{{K}}
\newcommand{\Kmct}{{\K}_\mathrm{MCT}}
\newcommand{\kernel}{\mathcal{C}}
\newcommand{\cd}{C_\mathrm{d}(t)}
\newcommand{\cpp}{C^{P}(t)}
\newcommand{\cfp}{C^{FP}(t)}
\newcommand{\cpf}{C^{PF}(t)}
\newcommand{\cff}{C^{F}(t)}
\graphicspath{{./figuras/}{./}}

\begin{document}
\title{On mean-field theories of dynamics in 
supercooled liquids}
\author{Marco Baity-Jesi}
\affiliation{Department of Chemistry, Columbia University, New York, NY 10027, USA}
\email[]{mb4399@columbia.edu}
\author{David R. Reichman}
\affiliation{Department of Chemistry, Columbia University, New York, NY 10027, USA}

\date{\today}

\begin{abstract}
We develop a hybrid numerical approach to extract the exact memory function $\K(t)$ of a tagged particle in three-dimensional glass-forming liquids. We compare the behavior of the exact memory kernel to two mean-field approaches, namely the standard mode-coupling theory and a recently proposed ansatz for the memory function that forms the basis of a new derivation of the exact form of $\K(t)$ for a fluid with short-ranged interactions in infinite dimensions.  Each of the mean-field functions qualitatively and quantitatively share traits with the exact $\K(t)$, although several important quantitative differences are manifest.
\end{abstract}

\maketitle

\section{Introduction}
Upon cooling, supercooled liquids (SCLs) display a dramatic increase in viscosity, the origin of which is still not fully understood~\cite{angell:95,berthier:11,charbonneau:17}. 
As the experimental glass transition is approached, dynamics become 
highly correlated and heterogeneous, making the construction of a first-principles microscopic theory a very challenging task. Historically, the first partially successful set of microscopically-based dynamical equations of SCLs was the mode-coupling theory (MCT)~\cite{gotze:08,reichman:05,janssen:18}, which takes as input only time-independent quantities
and outputs dynamical predictions via a mean-field-like factorization closure of the memory kernel $\K(t)$. 
The dynamical equations derived in this manner match experimental results in weakly supercooled liquids, but clearly deviate from them in the deeply supercooled regime.  In particular MCT predicts a power-law divergence of relaxation times at a temperature $T_\mathrm{MCT}$, which is much higher than the value $T_\mathrm{g}$ of the empirically measured glass transition temperature~\cite{fuchs:98}.

In the 1980s it was noticed that the dynamical equations for the MCT density-density correlation function have a form identical to that found for the spin correlation function in the $p=3$ $p$-spin model~\cite{kirkpatrick:87}, a paradigmatic member of a family of mean-field spin-glass models. 
This observation formed the basis for the random first-order theory (RFOT)~\cite{kirkpatrick:89}, which assumes that the equilibrium solution of the $p$-spin model, obtained within replica theory, also holds for supercooled liquids.
A direct implication of this connection is that MCT is a dynamical mean-field approximation which should become exact in the limit of infinite spatial dimensions, $d\to\infty$. This assumption was challenged in 2010, when it was observed numerically that with increasing dimensionality the replica and the MCT approaches give completely divergent transition points~\cite{ikeda:10,schmid:10}.

More recently, light has been shed on this discrepancy through a mean-field replica-based first-principles calculation showing that the RFOT scenario is the exact theory for the thermodynamics of high-dimensional sphere packings~\cite{charbonneau:14}. This was followed by a remarkable exact solution of the (mean-field) $d=\infty$ dynamics of simple SCLs~\cite{maimbourg:16}, which confirmed explicitly that the original MCT behaves incorrectly in the $d\to\infty$ limit.
It should be noted that the theory of Ref.~\cite{maimbourg:16}, which is also fully consistent with the dynamics of $p$-spin models, takes a completely different form than that of standard MCT. In particular, the mean-field theory of Ref.~\cite{maimbourg:16} forms a closed theory on the dynamical \emph{trajectories}, while MCT is closed in terms of correlation functions. Despite this fact, the exact $d=\infty$ solution does share many features with standard MCT.

In 2017 a transparent set of simple liquid state approximations was proposed which is capable of reproducing the correct $d\to\infty$ limit~\cite{szamel:17}. The main assumption, which holds in infinite dimensions, is that the memory function for tagged particle transport $\K(t)$ can be approximated through the real-space {\em diagonal} part of the pair force correlation function, which we denote $\cd$.
This approximation provides a direct link between the exact behavior of a fluid in $d=\infty$ (where this approximation becomes exact), and its three-dimensional counterpart. The validity of this approximation could be assessed by comparing $\K(t)$ and $\cd$ in $3d$ numerical simulations.
This is a challenging undertaking, however, because the memory function $\K(t)$ cannot be easily calculated directly from simulated dynamics. In particular, the time dependence of $\K(t)$ is generated with an unusual form of dynamics, and in the SCL regime
the behavior of the memory function spans several orders of magnitude in time. As a consequence, any route to the numerical evaluation of $\K(t)$ becomes a computationally daunting task.

In this paper, we use a combination of methods to accurately calculate the memory function $\K(t)$ in a three-dimensional glass-forming liquid. Several features of $\K(t)$ are compared with both $\cd$ and the more standard MCT approximation to $K(t)$ in three dimensions for the case of a canonical glass-forming liquid. We find clear quantitative differences between the exact $K(t)$ and these two functions. On the other hand, salient qualitative features are in agreement, implying that there are important features of liquids in $d=3$ that owe their properties to the mean-field, $d=\infty$ behavior.
Such insights pave the way for more sophisticated approximations in three dimensions that could potentially quantitatively capture the collective behavior of liquids near the experimental glass transition temperature.

In Sec.~\ref{sec:model}, we describe the system we simulate, and illustrate some important dynamical observables that result from our molecular dynamics simulations. In Sec.~\ref{sec:memory} we show how an exact self-consistent equation for the memory function $\K(t)$ can be derived through the projector operator technique, and we discuss the main assumptions of Ref.~\cite{szamel:17}. We then outline the methods used to calculate $\K(t)$ in Sec.~\ref{sec:method}, and present our results in Sec.~\ref{sec:results}. Finally, we summarize and discuss our findings in Sec.~\ref{sec:conc}.

\section{Model and simulations}\label{sec:model}
We simulate a three-dimensional Kob-Andersen 80:20 binary mixture~\cite{kob:94} of 1080 particles with molecular dynamics~\footnote{All the code used for this article is available at \texttt{github.com/mbaityje/STRUCTURAL-GLASS}.}, for which no finite-size effects in the autocorrelation times are expected or observed~\cite{berthier:12}.
We work in Lennard-Jones units~\cite{allen:89}, with
an integration time step of $dt=0.0025$, and a mass of all particles set to $m=1$.
The box is a periodic cube of linear size 
$L=(N/\rho)^{1/3}\approx9.65
$, chosen
so that the density $\rho=N/L^3=1.2$.  The interparticle potential at distance $r$ is
\begin{equation}\label{eq:V}
U_{ij}(r) = 4\epsilon_{ij}\left[\left(\frac{\sigma_{ij}}{r}\right)^{12}-\left(\frac{\sigma_{ij}}{r}\right)^{6}\right],\, 
\end{equation}
where $i,j\in\{\mathrm{A},\mathrm{B}\}$ indicate the particle type. The interaction energy amplitudes are $\epsilon_\mathrm{AA}=1, \epsilon_\mathrm{AB}=1.5$ and $\epsilon_\mathrm{BB}=0.5$, and the non-additive particle radii, chosen in order to suppress crystallization, are $\sigma_\mathrm{AA}=1, \sigma_\mathrm{AB}=0.8$ and $\sigma_\mathrm{BB}=0.88$. The potential is cut off and smoothed in order to have a continuous force at any distance~\footnote{
For our molecular dynamics simulations we have used the \texttt{hoomd} package, available at (\texttt{http://glotzerlab.engin.umich.edu/hoomd-blue}). The package is a \texttt{Python} API for GPU-based molecular dynamics of liquids, developed in Refs.~\cite{anderson:08,glaser:15}. The smoothing we applied follows the \texttt{xplor} mode described in the \texttt{hoomd} documentation. The smoothed potential is $U_{ij}^\mathrm{smooth}(r)=S(r)U_{ij}(r)$, where 
\begin{equation*}
S(r) =  \begin{cases}
         1,~~~~~ \text{if } r\leq r_\mathrm{on}\\
         \frac{(r_\mathrm{cut}^2-r^2)^2 (r_\mathrm{cut}^2+2r^2-3r_\mathrm{on}^2)}{(r_\mathrm{cut}^2-r_\mathrm{on}^2)^3}, ~ \text{if } r>r_\mathrm{on}>r_\mathrm{cut}\\
         0,~~~~~ \text{if } r\geq r_\mathrm{cut}\,,
        \end{cases}
\end{equation*}
where we used $r_\mathrm{on}=1.2\sigma$ and $r_\mathrm{cut}=2.5\sigma$.
The code used for these simulations is open-source and available at \texttt{www.github.io/mbaityje/STRUCTURAL-GLASS}.
}. For this model, there is an expected dynamical crossover at a temperature $T_\mathrm{d}\approx0.435$~\footnote{See Ref.~\cite{kob:94} for an estimation of the mode-coupling crossover.
The potential used in Ref.~\cite{kob:94} is shifted but not smoothed. This difference can induce a small shift in the value of the mode-coupling temperature in comparison to that of our system.}.
\begin{table}[bt]
\footnotesize
\begin{tabular}{c|cccccccccccc}
$T$ & 5.0 & 2.0 & 1.0 & 0.8 & 0.7 & 0.6 & 0.55 & 0.52 & 0.49 & 0.47 & 0.46 & 0.45\\\hline
$n_\mathrm{t}$ & 2000 & 1850 & 1750 & 1000 & 980 & 683 & 256 & 339 & 500 & 500 & 497 & 454\\
\end{tabular}
\caption{Temperatures used in our simulations. For each temperature, data was averaged over $n_\mathrm{t}$ trajectories.}
\label{tab:sim}
\end{table}
The temperatures $T$ employed in our simulations are given in Table~\ref{tab:sim}. For each $T$ we have thermalized 10 independent systems, and from each we run $n_\mathrm{traj}$ uncorrelated trajectories from which the autocorrelation functions detailed below are calculated. Observables are measured with a quasi-exponentially decreasing rate~\footnote{For each trajectory, we used an exponential succession of 1000 time points where measurements were taken. Since time steps are discrete, some times are duplicates and the grid results in having fewer points.}.

\begin{figure}[tb]
\includegraphics[width=\columnwidth, trim=0 70 0 0]{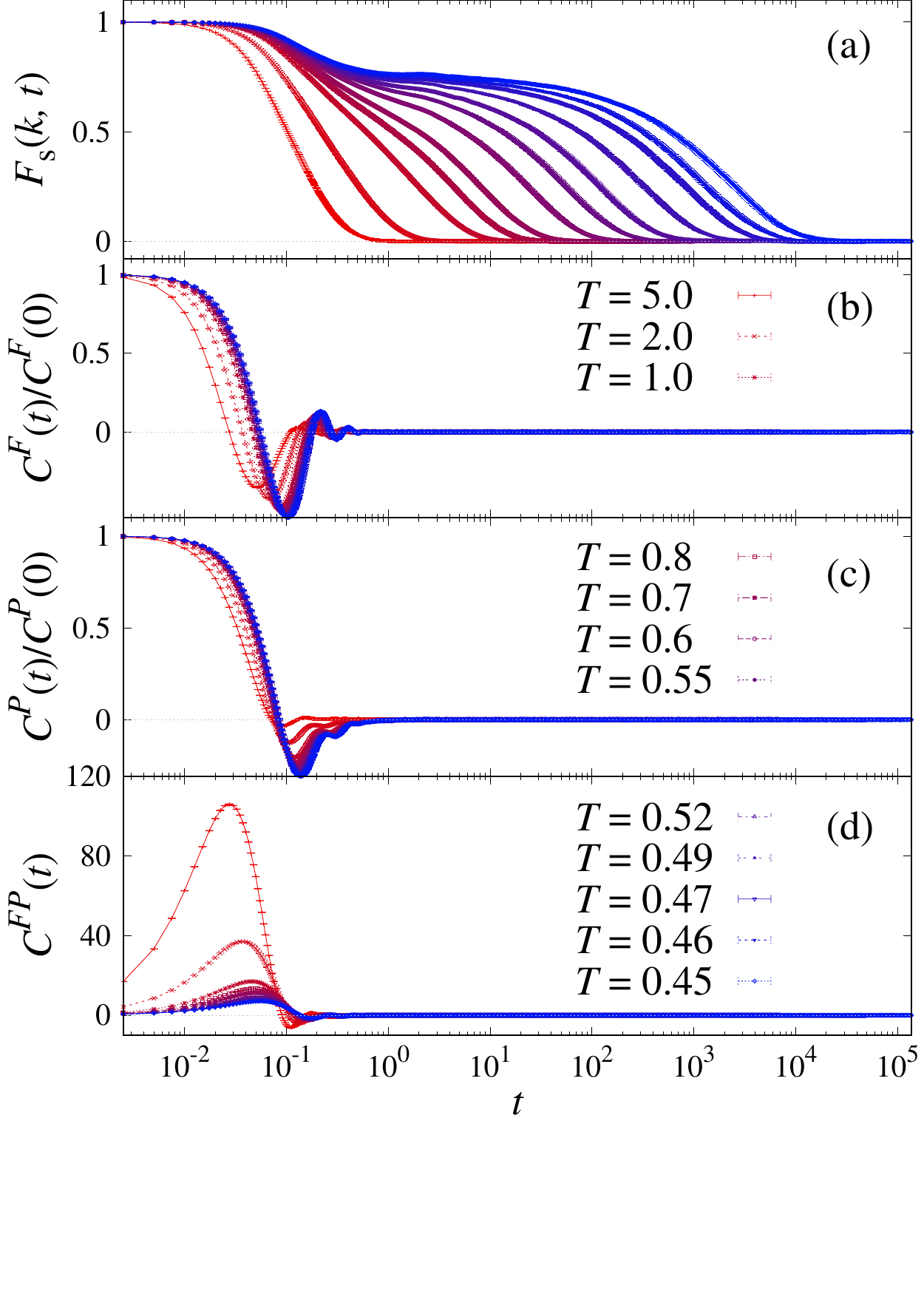}
\caption{
Autocorrelation functions at the temperatures depicted in Table~\ref{tab:sim}. Higher temperatures, in red, lead to correlation functions that decay faster; lower temperatures are in blue. The same color code is used throughout the article.
\textbf{(a)} Self-intermediate scattering function $F_\mathrm{s}(\vec k, t)$. We have averaged over the component permutations of wave vector $\vec{k}=\frac{2\pi}L(6,6,7)$.
\textbf{(b)} Momentum-momentum autocorrelation functions $\cpp$.
\textbf{(c)} Force-force autocorrelation functions $\cff$.
\textbf{(d)} Force-momentum autocorrelation functions $\cfp$.
}
\label{fig:trivial}
\end{figure}

In Fig.~\ref{fig:trivial}a we show the
self-intermediate scattering function 
$F_\mathrm{s}(\vec k,t)=\frac1N\langle\sum_{j=1}^N e^{i\vec{k}\cdot (\vec{r}_j(t)-\vec{r}_j(0))}\rangle$, 
with $\vec{k}$ denoting the wave vector, and $\vec{r}_j(t)$ the position of particle $j$ at time $t$.
Since here and throughout this paper all particles have mass $m=1$, particle velocities are equal to momenta, $P(t)$, and accelerations are equal to forces, $F(t)$. The autocorrelation functions of these observables, defined as
\begin{align}
 \cpp &= \left\langle P(0)P(t)\right\rangle\,,~~~~~~~\cff = \left\langle F(0)F(t)\right\rangle\,,\\[2ex]
 \cfp &= \left\langle F(0)P(t)\right\rangle =-\left\langle P(0)F(t)\right\rangle= -\cpf \,,
\end{align}
are shown in Fig.~\ref{fig:trivial}b-d. Note that $\dot{C}^{P}=\cpf$, and $\dot{C}^{FP}=\cff$, where we use the over-dot to indicate time derivatives.

\begin{figure}[t]
\includegraphics[width=\columnwidth]{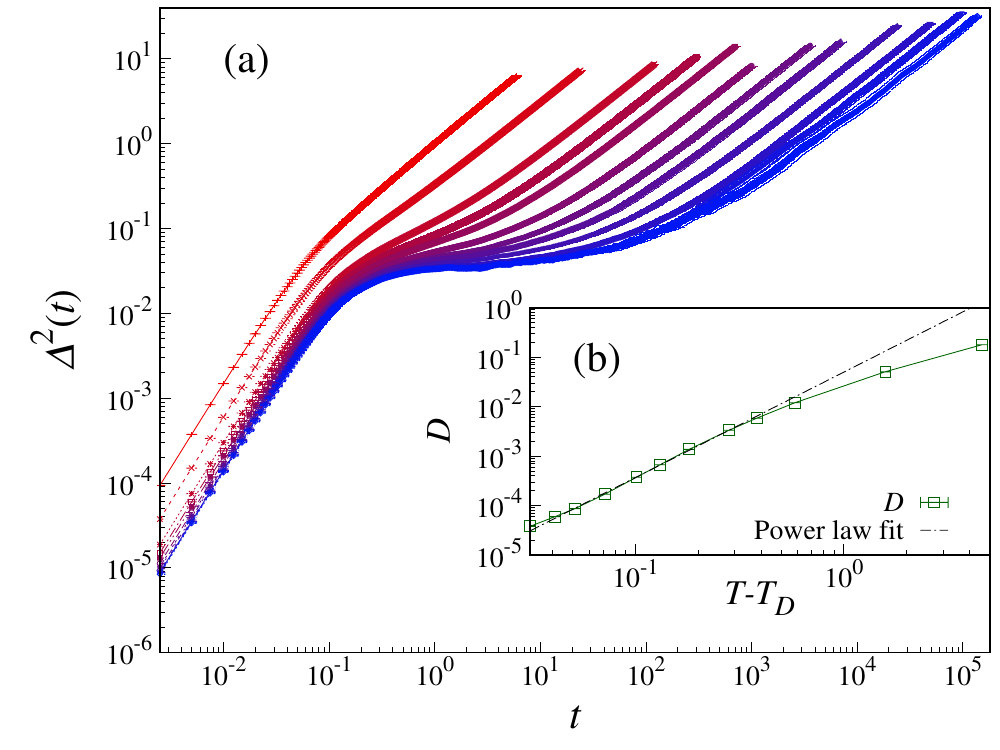}
\caption{
\textbf{(a)} Mean-square displacement $\varDelta^2(t)$ as a function of time, for all temperatures. Higher temperatures are leftwards, the color code is the same as in Fig~\ref{fig:trivial}.
\textbf{(b)} Diffusion coefficient $D$, as a function of $T-T_D$, on a logarithmic scale. $T_D=0.419(4)$ is obtained by fitting, for $T\leq0.7$, the vanishing of $D(T)$ through a function $f_D(T)=a_D (T-T_D)^{\eta_D}$. The value of the exponent is $\eta_D=2.13(7)$.
}
\label{fig:msd}
\end{figure}

In Fig.~\ref{fig:msd} we show the mean-square displacement of the particles, $\varDelta^2(t) = \left\langle\frac1N\sum_{i=1}^N\left[\vec{r}(t)-\vec{r}(0)\right]^2\right\rangle$, which for long times is related to the autocorrelation functions via $\varDelta^2(t) = 2t\int_0^t C^{P}(u)du \equiv 6tD$~\cite{balucani:95}. From the long-time behavior of $\varDelta^2(t)$, one can extract the diffusion coefficient, $D=\lim_{t\to\infty}\frac{\varDelta^2(t)}{6t}$~\footnote{
In the extrapolation, we took into account $\mathcal{O}(t^{-1})$ corrections to scaling. Thus, the fitting relation is $\frac{\varDelta^2(t)}{6t}=D+\kappa/t$, where $D$ and $\kappa$ are the fitting parameters.}. 
If we fit the vanishing of the diffusion coefficient to a power-law form $f_D(T)=a_D (T-T_D)^{\eta_D}$, with $T\leq0.7$, we obtain $T_D=0.419(4)$, $\eta_D=2.13(7)$.
These values should be taken \emph{cum grano salis}, since both exponent and $T_D$ depend on the fitting range~\cite{flenner:05}. However, the values we extract are grossly consistent with those previously reported in the literature~\cite{kob:94,flenner:05}.

\section{The tagged particle memory function}\label{sec:memory}
In a system defined by a Hamiltonian $H$, a generic observable $B(0)$ at time $t=0$ exhibits a formal time dependence for $t>0$
\begin{equation}
 B(t) = e^{i\mathcal{L}t}B(0)\,,~~~\dot{B}(t) = i\mathcal{L}B(t)\,,
\end{equation}
where the Liouville operator $\mathcal{L}$ is defined via 
\begin{equation}
 i\mathcal{L} = \sum_{i=1}^N\left[\frac{\partial H}{\partial P_i}\frac{\partial}{\partial Q_i}-\frac{\partial H}{\partial Q_i}\frac{\partial}{\partial P_i} \right]\,,
\end{equation}
where $P_i$ is the momentum of particle $i$, and $Q_i$ is its position.

The Mori-Zwanzig projection operator technique is based on projecting the dynamics onto a set of relevant observables (here, the impulse of a tagged particle), leaving the remaining degrees of freedom to evolve in an orthogonal dynamical subspace~\cite{mori:65,zwanzig:01,balucani:95}. We define the projection operator
\begin{equation}\label{eq:proj}
 \mathcal{P} B = \frac{\langle P(0) B \rangle}{\langle P(0)^2\rangle}P(0) = \frac{\langle P(0) B \rangle}{k_\mathrm{B}T}P(0)\,,
\end{equation}
where $P$ is one component of the impulse of a tagged particle. Throughout this work we set the Boltzmann constant to $k_\mathrm{B}=1$.
The orthogonal projector is defined as
\begin{equation}
 \mathcal{Q} = \mathbb{1}-\mathcal{P}\,.
\end{equation}

With the use of these projection operators the evolution of the system is split into two complementary sets of orthogonal dynamics, $i\mathcal{L}=i\mathcal{PL}+i\mathcal{QL}$, one projected onto the relevant variable(s) and one orthogonal to it (them).

One can show with the aid of these projectors that the momentum of a tagged particle follows the exact generalized Langevin equation~\cite{balucani:95}
\begin{equation}\label{eq:dotp}
 \dot{P}(t) = -\int_0^t \K(u) P(t-u) du + R(t)\,,
\end{equation}
where $R(t)=e^{i\mathcal{QL}t}\mathcal{Q}P(0)$ is usually called the fluctuating force, i.e. the force arising from the components of the system orthogonal to the relevant variable, which in this formalism may be thought of as a source of stochastic noise.
The function $\K(t)$ is the memory function, which takes the form of a force-force correlation function evolving with projected dynamics,
\begin{equation}\label{eq:Kdef}
 \K(t) = \frac{\left\langle F(0) e^{i\mathcal{QL}t} F(0) \right\rangle}{T}\,,
\end{equation}
where $F(0)$ is a component of the force acting on the tagged particle at time $t=0$.
$\K(t)$ is thus the time autocorrelation function of the fluctuating force.
Eq.~\eqref{eq:Kdef} is deceptively simple, since robust methods to compute exact \emph{projected dynamics} are scarce.

The definition of the projector in Eq.~\eqref{eq:proj} renders the fluctuating force $R(t)$ orthogonal to the momentum at all times, so Eq.~\eqref{eq:dotp} can be used to express the evolution of the momentum-momentum correlation function
in terms of the memory function as
\begin{equation}\label{eq:dotcp}
 \dot{C}^{P}(t) = - \int_0^t \K(t-u) C^{P}(u) du\,.
\end{equation}

The memory function $\K(t)$ can be written in a manner that does not involve orthogonal dynamics, by using in Eq.~\eqref{eq:Kdef} the Dyson relation~\cite{shi:03},
 \begin{equation}
 e^{i\mathcal{QL}t} = e^{i\mathcal{L}t}-\int_0^te^{i\mathcal{L}u}\mathcal{P}i\mathcal{L} e^{i\mathcal{QL}(t-u)}du\,, 
 \end{equation}
and by noticing that 
$\mathcal{P}(-i\mathcal{L})e^{i\mathcal{QL}(t-u)} F(0) = \langle F(0)e^{i\mathcal{QL}(t-u)}F(0)  \rangle P(0)/\langle P(0)^2\rangle$, which leads to an integral equation for $\K(t)$,
\begin{equation}\label{eq:Kselfc}
 \K(t) = \frac{1}{T}\left[\cff + \int_0^t C^{FP}(u) \K(t-u) \,du\right]\,.
\end{equation}
Eq.~\eqref{eq:Kselfc} has the advantage that 
the input to the integral equation, namely $\cff$ and $\cfp$, involve normal unprojected dynamics.
In principle, the solution of the integral equation~\eqref{eq:Kselfc}, generated with the exact input of $\cff$ and $\cfp$, provides an exact means to obtain $\K(t)$.

In Ref.~\cite{szamel:17} a physically-motivated set of approximations was proposed that provides a route to the the exact solution of the dynamics in $d=\infty$ for systems with short-ranged potentials~\cite{maimbourg:16}.
The principal assumption of Ref.~\cite{szamel:17} is that a mean-field theory for the memory function
can be obtained  
by discarding the components of the force-force autocorrelation function that involve more than two particles, and simultaneously replacing the projected dynamics with normal dynamics.
Explicitly, this means the replacement of $\K(t)$ with 
the diagonal force correlator,
\begin{equation}
 \cd = \frac{1}{3NT}\sum_{j=1}^N \sum_{i>j}^N\sum_{\alpha=x,y,z}\left\langle F_{ji,\alpha}(0)F_{ji,\alpha}(t) \right\rangle
\end{equation}
where we choose the tagged particle to be particle 1, and $F_{ji,\alpha}(t)=e^{i\mathcal{L}t}F_{ji,\alpha}(0)~[\alpha=x,y,z]$ is a component of the force that particle $i$ exerts on particle $j$ at time $t$.
Clearly, $\cd$ is the force-force correlation $\cff$ where all the non-diagonal terms which couple different particles at different times are discarded. This function is expected to decay slower than $\cff$, due to cancellations between the diagonal and non-diagonal components of $\cff$~\cite{vergeles:99}.

While the substitution proposed in Ref.~\cite{szamel:17},
\begin{equation}\label{eq:KCd}
\K(t)\to\cd \,,
\end{equation}
is a key step for the recovery of the \emph{exact} dynamics in $d=\infty$, it may be viewed as an interesting, albeit uncontrolled, mean-field-like approximation in $d=3$. It should be noted that this mean-field relationship leads to a picture distinct from that of the canonical mean-field theory of supercooled liquids, namely MCT, which is based on an uncontrolled factorization of density modes in $k$-space~\cite{gotze:08}. 

To calculate the tagged particle memory function within canonical MCT, one can project the forces in Eq.~\eqref{eq:Kdef} on the pair density modes, which are expected to give the dominant contribution to the tagged particle dynamics~\cite{balucani:95}.
In practice, this amounts to projecting the forces onto the wave-vector dependent solute self-density and solvent collective density, $\hat b(\vec k)=\sum_{i=2}^Ne^{i\vec k\cdot(\vec r_i-\vec r_1)}$, through the projection operator~\cite{balucani:95,egorov:02}
\begin{equation}\label{eq:P2}
 \mathcal{P}_2 (B)= \sum_{\vec k}\frac{\left\langle \hat b(-\vec k) B\right\rangle}{N S(\vec k)}\hat b(\vec k)\,,
\end{equation}
where $S(\vec k)=\langle\frac1N\sum_{i,j=1}^Ne^{i\vec k\cdot(\vec r_i-\vec r_j)}\rangle$ is the static structure factor.
The MCT memory function involving these modes is then expressed as
\begin{equation}\label{eq:KmctDef}
 \Kmct(t)=\frac{\left\langle [\mathcal{P}_2 F(0)] e^{i\mathcal{LQ}t}  [\mathcal{P}_2 F(0)]\right\rangle}{T}\,,
\end{equation}
where $F(0)$ is the $\alpha^\mathrm{th}$ component of the force at time $t=0$.
By replacing Eq.~\eqref{eq:P2} into Eq.~\eqref{eq:KmctDef} one obtains~\cite{balucani:95,egorov:02}
\begin{equation}\label{eq:Kmct-int}
\Kmct(t) = \frac{T}{8\pi^3\rho} \int d\vec k\, k_\alpha^2 \,c(\vec k)^2 F_\mathrm{s}(\vec k,t)F_\mathrm{c}(\vec k,t)
\end{equation}
where $F_\mathrm{c}(\vec k,t) = \langle
 \sum_{j,\ell=1}^N e^{i\vec k\cdot (\vec r_j(0)-\vec r_\ell(t))}\rangle$ is the collective scattering function
and $c(\vec k)=\frac1\rho\left(1-\frac1{S(\vec k)}\right)$.
It is important to remark that the short-time behavior of $\K(t)$ cannot be captured by Eq.\eqref{eq:Kmct-int}. In particular, an integration over all $k$ will result in the near complete suppression of the slow dynamical regime. The usual procedure that is followed is that the short time kinetic behavior is subtracted and reintroduced phenomenologically via the addition of a function such as that in Eq.~\eqref{eq:cshort}~\cite{balucani:95}. However, we find that such a procedure is not consistent since the true amplitude of $\K(0)$ is significantly smaller than that of $\Kmct(0)$. In the following, we choose a cutoff value of $k$, $k_\mathrm{max}=28$, for which the long-time behavior of $\Kmct(t)$ is converged and stable and for which a reasonable description of the short time behavior is captured.

Our goal for the remainder of this work will be to compare $\cd$ and $\Kmct(t)$ to the exact $\K(t)$ as extracted from molecular dynamics simulations. We note that our comparisons will differ in a subtle manner from a full comparison of either the mean-field theory of Ref.~\cite{maimbourg:16} or of MCT~\cite{gotze:08} to the exact behavior of $\K(t)$ in $d=3$.
In particular, here we use the \emph{exact} molecular dynamics in $d=3$ to evaluate the respective mean-field expressions as opposed to solving the non-linear mean-field equations self-consistently. Thus, the comparisons we make will deviate somewhat from those associated with complete theories as outlined in Refs.~\cite{maimbourg:16,szamel:17} and Ref.~\cite{voigtmann:04}, respectively.
We will return to this point later in our work and in the conclusions. This \emph{caveat} aside, we do expect the comparisons to be revealing of the gross successes and failures of the various mean-field approaches in low spatial dimensions.

\section{Numerical evaluation of the memory function}\label{sec:method}

A numerical evaluation of the tagged particle memory function $\K(t)$ is challenging due to the fact that the dynamical evolution of the force is generated by projected dynamics. We have found that recent proposals to directly generate a molecular dynamics for projected dynamics~\cite{carof:14} are successful at high temperatures where the memory function is simple in structure and decays quickly in time, but the resulting equations become very stiff in dense fluids at low temperatures, rendering the procedure unstable.

 To deal with the difficult task of evaluating the memory function over the full span of distinct dynamical regimes, we have adopted a piecewise concatenation of two separate methods, one of which is accurate at long times, the other of which is accurate at short times.
For the evaluation of the long-time behavior of $\K(t)$, we Laplace transform Eq.~\eqref{eq:dotcp}, obtaining
\begin{equation}\label{eq:Ks}
 \hat\K(s) = \frac{T-s\,\hat{C}^{P}(s)}{\hat{C}^{P}(s)}\,,
\end{equation}
where $\hat{A}(s)$ indicates the Laplace transform of the function $A(t)$, and we notice that $\hat{C}^P(0)=\int_0^\infty\langle P(0)P(t)\rangle=T$.
We then  invert the Laplace transform with the Gaver-Stehfest method~\cite{gaver:66,stehfest:70}. This procedure, which we will call the Laplace method in the following, is capable of estimating $\K(t)$ at long times, although it requires averaging over a large number of trajectories. In addition there are a series of {caveats} related to the precision of the approach which we discuss in Appendix~\ref{app:laplace}.

For the short-time behavior of $\K(t)$, we note that Eq.~\eqref{eq:Kselfc} is a Volterra integral equation of the second kind. The integral on the right-hand side can be decomposed with a trapezoidal integration scheme, and solved in $\mathcal{O}(t^2)$, by passing the last element of the integral to the left-hand side of Eq.~\eqref{eq:Kselfc}~\cite{press:07}. This algorithm, which we will call the Volterra method, is very accurate for short times. Details are given in Appendix~\ref{app:volterra}.

The memory functions $\K(t)$ that we report are a piecewise concatenation of the $\K(t)$ obtained through the Volterra method for $t\lesssim0.05$, and the Laplace method for $t\gtrsim0.05$. As shown in Appendix~\ref{app:checks}, the two methods give similar solutions around $t=0.05$, and the resulting $\K(t)$ describes appropriately the dynamics, satisfying the consistency check of Sec.~\ref{sec:checks} below. Due to the instability of the inverse Laplace transform (see Appendix~\ref{app:laplace}), we had to average our data over a large number $n_t$ of trajectories (see Tab.~\ref{tab:sim}).

\subsection{Consistency check on the obtained memory functions}\label{sec:checks}
To ensure that the obtained $\K(t)$ is actually a faithful representation of the memory function $\K(t)$, we calculate $\cfp$ via Eq.~\eqref{eq:dotcp} after obtaining $\K(t)$ with the approach outlined above, and we compare it with the one we measured directly from molecular dynamics. We then integrate the obtained $\cfp$ over time with a trapezoidal integration scheme, and compare it to the measured $\cpp$, since $\cpp$ contains more visually recognizable features than does $\cfp$. In Appendix~\ref{app:checks} we show the results of this consistency check.

\section{Results}\label{sec:results}
\begin{figure}[tb]
\includegraphics[width=\columnwidth, trim=110 0 0 0]{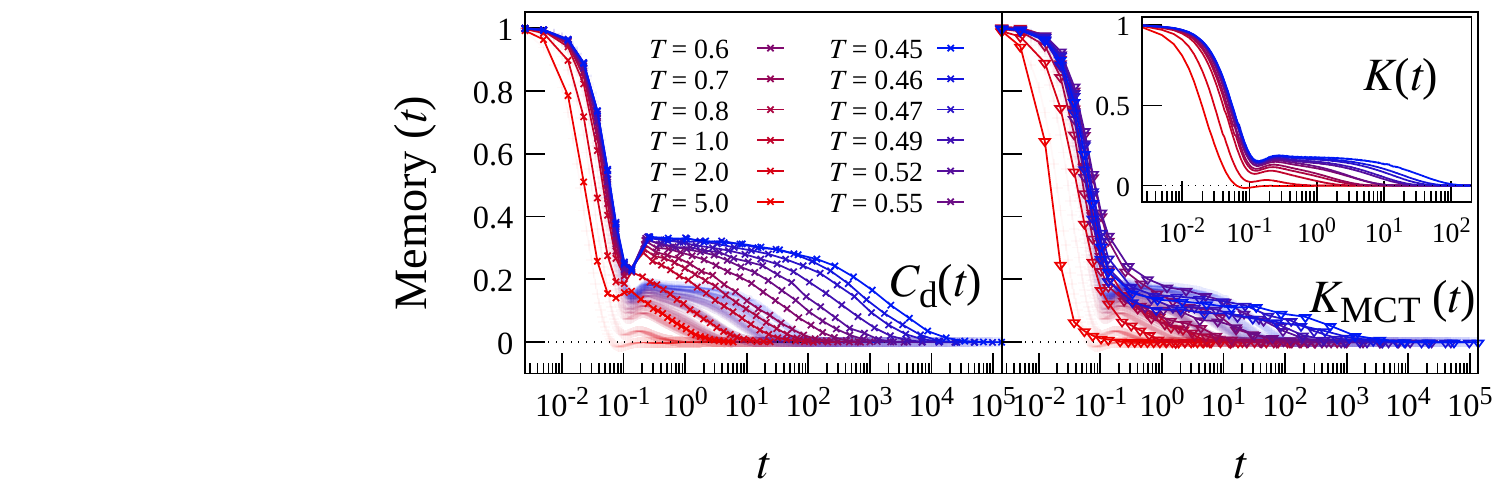}
\caption{\textbf{Left:} Normalized diagonal autocorrelation $\cd$ (solid $\times$ line-points) for all the simulated temperatures.
\textbf{Right:} MCT memory function $\Kmct(t)$ (solid $\triangledown$ line-points) 
for all the simulated temperatures.
$\K(t)$ is shown in transparency in both plots, as a reference for comparison.
\textbf{Inset:} Memory function $\K(t)$.
Error bars are not shown for visual clarity, see Appendix~\ref{app:checks} for a discussion of the error bars.}
\label{fig:KCd}
\end{figure}

\begin{figure}[tb]
\includegraphics[width=\columnwidth,trim=0 0 15 0]{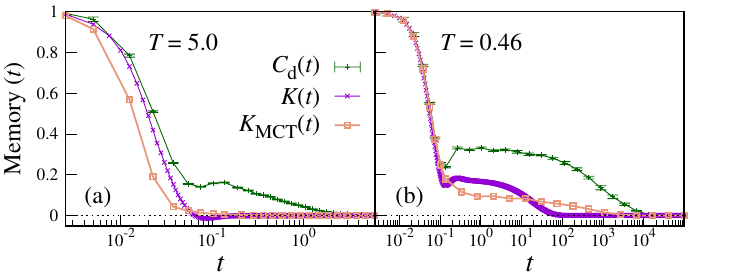}
\caption{Normalized MCT memory function $\Kmct(t)$, compared to the 
exact memory function $\K(t)$ and the diagonal correlation function $\cd$. \textbf{(a)} $T=5.0$. \textbf{(b)} $T=0.46$. 
}
\label{fig:Kmct}
\end{figure}

In Fig.~\ref{fig:KCd} we show the tagged particle memory function $\K(t)$, the diagonal force autocorrelation function $\cd$ and the memory function calculated with a traditional MCT-type theory, $\Kmct(t)$, at different temperatures. 
For better visual clarity, in Fig.~\ref{fig:Kmct} we show the three functions at two extreme temperatures.
As discussed in Sec.~\ref{sec:memory}, the expression of Eq.~\eqref{eq:Kmct-int} only provides a description of the long-time behavior of $\K(t)$. In particular, neither the rapid decay of $\K(t)$ nor the relative amplitude of the plateau height compared to $K(0)$ may be obtained from it. On the other hand, $\cd$ does provide an approximate and consistent description in both regimes.
In a qualitative sense, the shape of the plateau region of $\K(t)$ is also more accurately described by $\cd$ than by $\Kmct(t)$, including the stability of the plateau and the dip in $\K(t)$ close to $t=0.1$, which separates the short and long time regimes.

\begin{figure}[!tb]
\includegraphics[width=\columnwidth, trim=0 70 0 0]{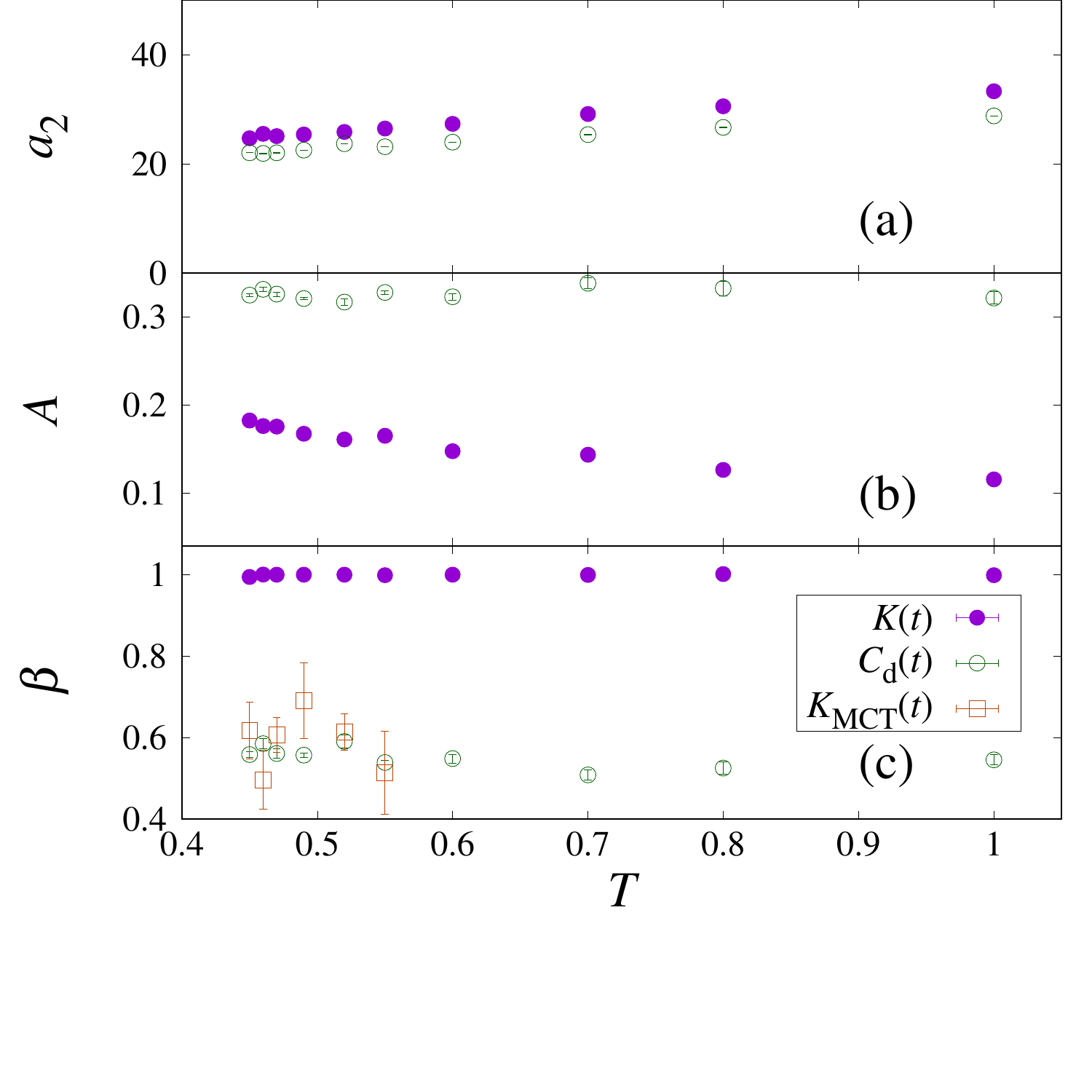}
\caption{
\textbf{(a)}
Fitting parameter $a_2$ [Eq.~\eqref{eq:cshort}] for the short-time behavior of $\K(t)$ and $\cd$. 
\textbf{(b)}
Height of the plateau $A$ as a function of temperature.
\textbf{(c)}
Stretching exponent $\beta$. The stretching exponent of $\Kmct(t)$ is also shown for $T<0.6$ (higher temperatures are not shown because of large error bars).
$A$ and $\beta$ are obtained by fitting the long-time decay of the autocorrelation functions to Eq.~\eqref{eq:clong}.
}
\label{fig:short}
\end{figure}

As is clear from Fig.~\ref{fig:KCd} and as noted above, the short-time behaviors of $\K(t)$ and $\cd$ are quantitatively similar, both in terms of the their respective rapid decreases as well as the appearance of an interesting (almost temperature-independent) oscillation that separates the short and long time regimes. 
To quantify the similarities between $\K(t)$ and $\cd$, we fit the short-time behavior to~\cite{balucani:95,kob:02} 
\begin{equation}\label{eq:cshort}
C_\mathrm{short}(t)=\frac{a_{1}}{\cosh(a_2 t)}\,, 
\end{equation}
where $a_1=1$ when the functions are normalized, so the only free parameter is $a_2(T)$, which we show in Fig.~\ref{fig:short}a.
Clearly, even the mild temperature dependence of the rapid initial decay of $\K(t)$ is captured bu the mean-field diagonal approximation $\cd$.

The long-time decay of the memory functions can be fit to a stretched exponential form,
\begin{equation}\label{eq:clong}
 C_\mathrm{long}(t) = A\exp[(t/\tau)^\beta]\,,
\end{equation}
where the plateau height $A$, the autocorrelation time $\tau$, and the stretching exponent $\beta$ are free parameters.

In Fig.~\ref{fig:short}b we see that, although the plateau height for $\cd$ is larger than for $\K(t)$, the former does not depend on temperature, whereas the latter grows mildly as $T$ decreases. 
The stretching exponent $\beta$ (Fig.~\ref{fig:short}c) is different for the two functions: $\K(t)$ has exponential decay ($\beta=1$), whereas $\cd$ has a stretched exponential decay, with $\beta\approx0.55$. 
It should be noted that the same stretched exponential behavior emerges in $\Kmct(t)$.
The possibility exists that, even though our numerical procedure appears to be converged and passes non-trivial consistency tests, that it is still not capable of describing accurately the tail of $\K(t)$. Future work will be devoted to this important issue.
\begin{figure}[!tb]
\includegraphics[width=0.95\columnwidth]{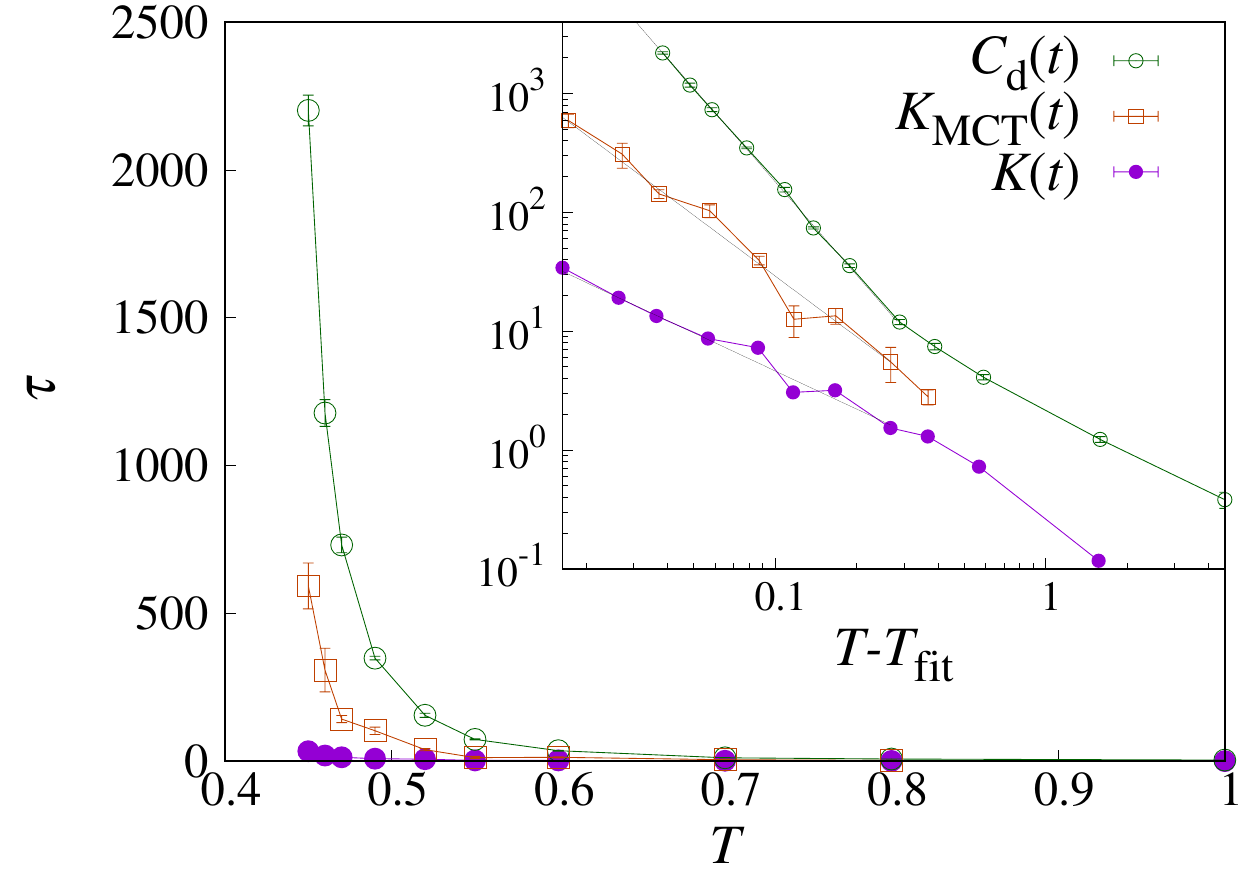}
\caption{Autocorrelation time $\tau(T)$
calculated from the long-time behavior, Eq.~\eqref{eq:clong}, of $\K(t)$, $\cd$ and $\Kmct(t)$. The inset shows the same quantity as a function of $T-T_\mathrm{fit}$, on a logarithmic scale. Note that $T_\mathrm{fit}$ is not the same for the three functions (see main text).
}
\label{fig:long}
\end{figure}

The autocorrelation times $\tau$ of the memory functions are shown in Fig.~\ref{fig:long}. Even though the autocorrelation times for $\cd$ are much larger at the same temperature than those of $\K(t)$, both exhibit a form consistent with a power law growth $\tau\propto (T-T_\mathrm{fit})^{-\eta_\mathrm{fit}}$ which extrapolates to similar dynamical crossover temperatures $T_\mathrm{fit}$, but with different exponents $\eta_\mathrm{fit}$. We also include the autocorrelation times extracted from long-time fits of $\Kmct(t)$.
When fitting in the range $T\leq0.7$, extrapolated dynamical temperatures derived from the exact memory function, the diagonal force autocorrelation and the MCT memory function are, respectively, $T_\mathrm{fit}^\mathrm{(mem)} = 0.434(17)$, $T_\mathrm{fit}^\mathrm{(diag)} = 0.412(3)$ and
$T_\mathrm{fit}^\mathrm{(MCT)} = 0.433(8)$. 
The exponents of the power-law growth are $\eta_\mathrm{fit}^\mathrm{(mem)} = 1.1(3)$,  $\eta_\mathrm{fit}^\mathrm{(diag)} = 2.61(9)$ and $\eta_\mathrm{fit}^\mathrm{(MCT)} = 1.7(3)$.
Clearly the exact and approximate memory functions qualitatively exhibit similar forms of decay. However, the diagonal and MCT approximations produce a much more rapid growth of relaxation times.

\begin{figure}[!tb]
\includegraphics[width=\columnwidth]{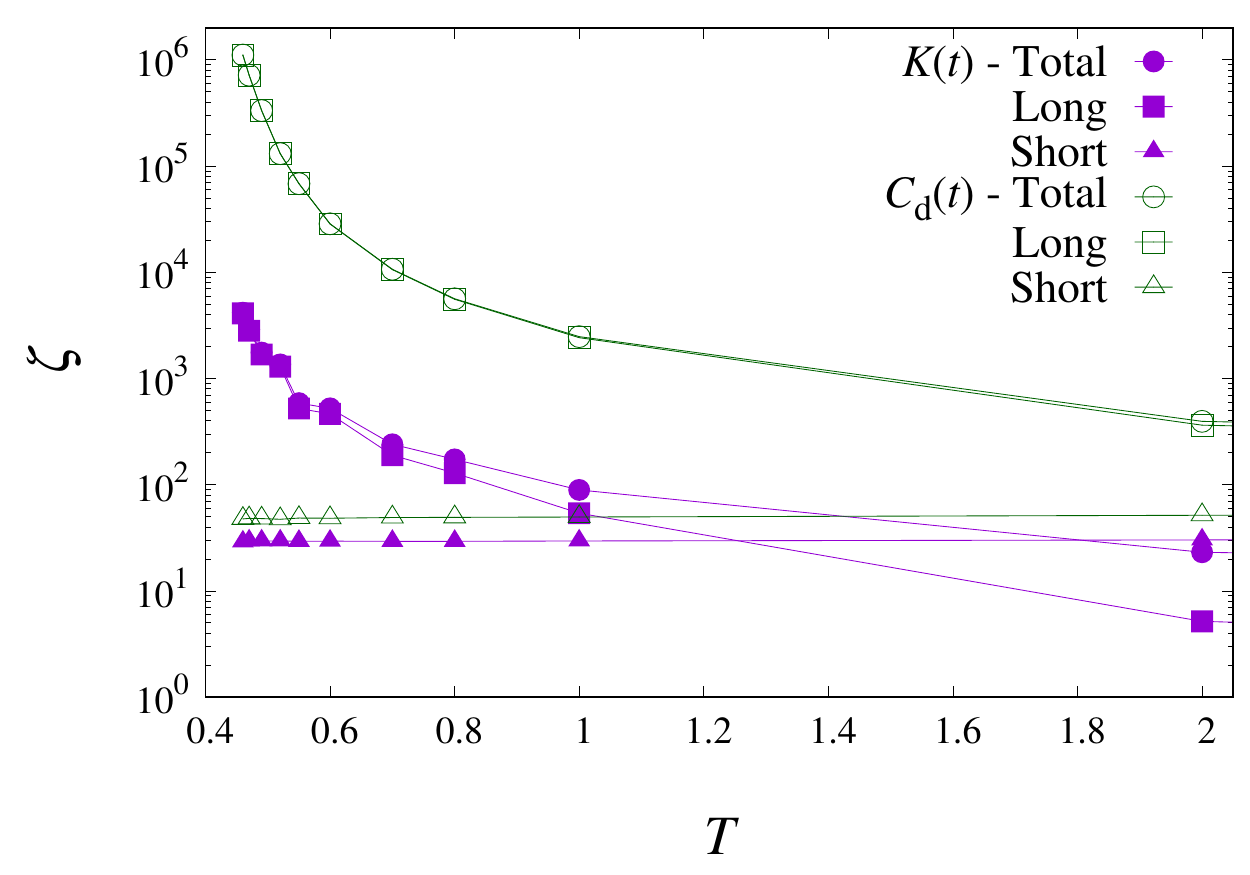}
\caption{Friction coefficient $\zeta$ as a function of temperature $T$. 
The purple full points represent the friction coefficients stemming from the memory function $\K(t)$, whereas the green empty points indicate the friction measured from $\cd$.
Circles are the $\zeta$ calculated on the full correlation functions [Eq.~\eqref{eq:friction}], triangles are the friction coefficients resulting by the integration of only the short-time part of the autocorrelation functions, $C_\mathrm{short}(t)$, derived from Eq.~\eqref{eq:cshort}. 
Squares result from the integration of the long-time correlation function: $C(t)-C_\mathrm{short}(t)$.
}
\label{fig:friction}
\end{figure}

A further measure of comparison between exact and approximate memory functions is obtained via the friction coefficient
\begin{align}\label{eq:friction}
\zeta^\mathrm{(mem)} &=\rho\int_0^\infty \K(t)\,dt\,, \\
\zeta^\mathrm{(diag)}&=\rho\int_0^\infty \cd  \,dt\,,
\end{align}
which we show in Fig.~\ref{fig:friction}.
Power law fits, for $T\leq0.7$, of the form $\zeta(T)=B_\zeta(T-T_\zeta)^{-\eta_\zeta}$ show a similar scenario to the one found bt fitting the long-time behavior of $\K(t)$ and $\cd$, with compatible $T_\zeta$ but different $\eta_\zeta$. The extrapolated dynamical temperatures are  $T_\zeta^\mathrm{(mem)} = 0.43(2)$ for the memory function, and $T_\zeta^\mathrm{(diag)} = 0.414(6)$ for the diagonal function, and the exponents are $\eta_\zeta^\mathrm{noise}=1.4(3)$ and $\eta_\zeta^\mathrm{diag}=2.6(2)$. 
To rule out the possibility that our extrapolations are biased by preasymptotic effects arising from the nature of the short-time behavior of the memory functions which barely depends on temperature, we have recalculated the friction coefficients by removing from $\K(t)$ and $\cd$ the short-time part of the decays fitted via Eq.~\eqref{eq:cshort}. 
After this procedure, the qualitative situation remains unchanged, with the dynamical temperatures being
$T_\zeta^\mathrm{mem,long} = 0.42(2)$ and $T_\zeta^\mathrm{diag,long} = 0.414(6)$, and the exponents $\eta_\zeta^\mathrm{noise, long}=1.6(4)$ and $\eta_\zeta^\mathrm{diag,long}=2.6(2)$.
This can be understood by remarking that the plateaus of the normalized memory functions in Fig.~\ref{fig:KCd} are at $\sim\mathcal{O}(1)$ height, whereas the span of the short-time regime is a tiny fraction of the total time.

\section{Discussion}\label{sec:conc}

In this paper we have devised a numerically exact scheme to reconstruct the memory function for tagged particle motion in SCLs.  The approach relies on the combination of the solution of a Volterra integral equation with numerically exact input and the numerical inversion of the solution for the memory function in Laplace space using the Gaver-Stehfest approach.  The former method is accurate at short times while the latter is accurate at long times.  The success of the concatenation of these approaches, which agree on intermediate time scales, is indicated via the reproduction of simulated transport data.  Our approach thus opens the door for the detailed assessment of theories of SCLs that utilize the memory function approach.

We have compared the temperature dependence of the memory function extracted from molecular dynamics simulations to two distinct mean-field approaches. In particular, we consider the standard MCT tagged particle memory function filtered with exact structural and dynamical input as well as the unprojected diagonal force-force correlation function as approximate forms for the tagged particle memory function.  The latter quantity is related to the exact memory function for a fluid interacting with short ranged forces in infinite dimensions.  We find that the approximate approaches share qualitative (and in some cases quantitative) features with the exact memory kernel in three dimensions while also exhibiting important differences.  Focusing on the diagonal force-force correlation function, we find that the short time behavior of the memory function is quantitatively reproduced, as is the extrapolated temperature of a putative power-law singularity of the long time relaxation.  On  the other hand the exponent of the growth of relaxation times as well as the form of the long-time relaxation of these two functions (stretched exponential versus exponential) differ markedly.
This distinction highlights the importance of carrying out a fully self-consistent calculation of the mean-field dynamics of Ref.~\cite{maimbourg:16} as opposed to using exact non mean-field trajectories within the framework of the mean-field function $\cd$. 
The differing rate of growth of the relaxation could be attributed to this difference in the calculation procedure. 
On the other hand, some discrepancies, such as those associated with the stretching exponent $\beta$, might arise from the nature of our numerical determination of $\K(t)$.

Our observations provide useful information concerning the ability of approximate mean-field approaches to capture the realistic behavior of particle motion in three-dimensional SCLs, as well as guidelines on the construction of more sophisticated theories that build on mean-field foundations.  It should be noted that the comparisons made here do not employ fully self-consistent solutions of the mean-field dynamics, and instead use the exact three-dimensional trajectories in conjunction with mean-field expressions.  It is possible that this inconsistency degrades the level of agreement with the exact memory kernel.  Future work will be devoted to this issue.  In addition, we will compare the exact memory function as presented here with advanced approaches that numerically bridge the behavior of the infinite dimensional and three dimensional limits such as ``cluster dynamical mean-field" theories~\cite{biroli:19} based on Ref.~\cite{szamel:17}.

\begin{acknowledgments}
We thank G. Szamel, P. Charbonneau, I. Dunn, B. Kloss, F.P. Landes, A. Manacorda and F. Zamponi for inspiring discussions.
We thank Ian Dunn for providing his Prony function fitting package, and F.P. Landes for useful interactions about molecular dynamics simulations. 
This work was funded by the Simons Foundation for
the collaboration “Cracking the Glass Problem” (No. 454951 to D.R. Reichman). 
This work benefited from access to the University of Oregon high performance computer, Talapas.
\end{acknowledgments}

\appendix

\section{Computing the memory function through the Laplace method}\label{app:laplace}
The Laplace transform of a function $C(t)$ is
\begin{equation}\label{eq:laplace-transform}
 \hat{C}(s) = \int_0^\infty e^{-st} C(t)\, dt\,,
\end{equation}
where $s\in\mathbb{C}$ is a complex number. Eq.~\eqref{eq:laplace-transform} can be used to pass from Eq.~\eqref{eq:dotcp} to Eq.~\eqref{eq:Ks} without any complication, because $\cpp$ is locally integrable.

Eq.~\eqref{eq:Ks} needs then to be inverted. The inverse Laplace transform, is defined by the Bromwich integral~\cite{josso:12}
\begin{equation}
 C(t) = \frac{1}{2\pi i} \lim_{\delta\to\infty} \int_{\gamma-i\delta}^{\gamma+i\delta} e^{st} \hat{C}(s)ds\,,
\end{equation}
where $\gamma$ is a real number such that it is larger than the real part of all singularities.

To calculate the inverse transform we use the Gaver-Stehfest algorithm~\cite{gaver:66, stehfest:70,spendier:10}, which has the additional benefit of giving a simple understanding of the errors and origins of possible instabilities~\cite{josso:12,kuznetsov:13}.

The Gaver-Stehfest inversion algorithm approximates $C(t)$ through a finite series of functions 
\begin{equation}
 C_n(t) = \frac{\log(2)}{t}\sum_{k=1}^{2M} \omega_k\hat{C}\left(\frac{k\log(2)}{t} \right)\,,
\end{equation}
where the coefficients are given
\begin{align}
\omega_k&=\\ 
&(-1)^{M+k} \sum_{j=\lfloor\frac{k+1}{2}\rfloor}^{\min(k,M)} \frac{j^{M} (2j)!}{(M-j)!\, j!\,(j-1)!\,(k-j)!\,(2j-k)!}\,, \notag
\end{align}
with $\lfloor x \rfloor$ denoting the integer part of $x$. 
The values of $\omega_k$ can be calculated first, and stored in a hash table.
In practice, the Laplace transform only needs to be evaluated for real values of $s$~\footnote{To evaluate the functions at points off the grid of measurements, we fitted the data through a sum of exponentials using 
Beylkin and Monz\'on's approximate Prony method~\cite{beylkin:05}.
The implementation we adopted was written for Ref.~\cite{dunn:19}, and is publicly available at \texttt{github.com/iansdunn/prony}.
}, which we call $p\in\mathbb{R}$.

\paragraph*{Noise and precision}
The Gaver-Stehfest algorithm depends on a sum of $2M$ terms. The larger $M$, the more accurate the inversion. The drawback is that large-$k$ terms in this sum amplify small fluctuations, so $M$ cannot be too large because otherwise the noise is amplified, resulting in a huge loss in precision. Instructive benchmarks on the optimal $M$ are performed in Ref.~\cite{josso:12}.
A general rule of thumb is that $2M$ should correspond to the decimal digit where the errors appear, so if the only source of noise is the rounding error, then $M=7$ is the best choice in computations with double precision accuracy.

In our case, there are many sources of noise and systematic error. 
Even for exact functions, the fact that one interpolates between points on a grid is, \emph{per se}, a source of error.
In addition, our data is intrinsically noisy due to the fact that it arises from numerical simulations that must be averaged. Lastly, given the disparate time scales involved in the decay of $\K(t)$, data needed to be collected on a (quasi) exponentially spaced time grid, implying that at large $t$ a single measurement is representative of a very large time span.

In this work we used $M=5$, which allowed for stable inverse transforms with our data at every temperature (although we needed a large number of trajectories, see Tab.~\ref{tab:sim}). A low value of $M$ implies a loss in accuracy on fine-grained scales, which we show has no dramatic consequence on the reconstruction of the memory kernel (see Appendix~\ref{app:M}).

\paragraph*{Integration ranges}
Another issue is with the Gaver-Stehfest method is that it requires the evaluation of the Laplace transforms at very large values of $p$, with 
\begin{equation}
p_\mathrm{max}=\frac{2M\log(2)}{t_\mathrm{min}} \,,
\end{equation}
where $p_\mathrm{max}$ is the largest value of $p$, and $t_\mathrm{min}=dt=0.0025$ is the smallest time. 
When $p$ is too large, the integrand in~\eqref{eq:laplace-transform} incurs numerical underflow.
This implies again that $M$ cannot be too large, and that time cannot be too small. As a consequence, the Laplace inversion becomes extremely noisy at small $t$, which is why concatenation with a method suitable at short times is required.

\section{Computing the memory function through the Volterra integration}\label{app:volterra}
Here, we follow Ref.~\cite{press:07}, with minor adaptations to our problem.
Eq.~\eqref{eq:Kselfc} is a Volterra equation of the second kind.
By defining $t_j=jdt~(j\in\mathbb{N}^+)$, and $A_j=A(t_j)$, we can rewrite it using a trapezoidal integration scheme on a linear grid
\begin{equation}\label{eq:Kselfc-discrete}
 \K_i = \frac{C^{F}_i}{T} + \frac{dt}{2T}\left( C^{FP}_{i0}\K_0 + 2\sum_{j=1}^{i-1}C^{FP}_{ij}\K_j + C^{FP}_{ii}\K_i\right)\,.
\end{equation}
Since $C^{FP}(t,u)=C^{FP}(t-u)$, then $C^{FP}_{ii}=0$, thus Eq.~\eqref{eq:Kselfc-discrete} becomes
\begin{equation}\label{eq:Kselfc-discrete-short}
 \K_i = \frac{C^{F}_i}{T} + \frac{dt}{2T}\left( C^{FP}_{i0}\K_0 + 2\sum_{j=1}^{i-1}\kernel_{ij}\K_j\right)\,.
\end{equation}

Since $C^{FP}(t)$ decays to zero, at large times, its signal-to-noise ratio is low. 
In order to improve accuracy, we can define a $t^*=i^*dt$ such that $\kernel(t)\approx0~\forall t>t^*$, and $\kernel(t-u)\approx0~\forall u<t-t^*$. Similar procedures are used to reduce the statistical error in spin-glass correlation functions~\cite{janus:09b}, with $t^*$ the time when the signal-to-noise ratio becomes smaller than an order 1 number. 
Since the grid is linear, $\kernel_{ij}\approx0 ~\forall j<i-i^*$, so Eq.~\eqref{eq:Kselfc-discrete-short} further simplifies to
\begin{equation}\label{eq:Kselfc-discrete-short-star}
 \K_i = \frac{C^{F}_i}{T} + \frac{dt}{2T}\left(\ C^{FP}_{i0}\K_0 + 2\sum_{j=\max(0,i-i^*)}^{i-1}C^{FP}_{ij}\K_j\right)\,.
\end{equation}

\paragraph*{Generic grid}
If the time grid is non-linear, as in our case, the discretized Volterra equation is 
\begin{align}\label{eq:Kselfc-disc-nonl}
 \K_i =& ~\frac{C^{F}_i}{T} +\\ \notag
 +& \frac{1}{2T}\bigg[\sum_{j=1}^{i-1}
 \left(
 C^{FP}_{ij}\K_j + C^{FP}_{i{j-1}}\K_{j-1}
 \right)
 \left(
 t_j-t_{j-1}
 \right)\\\notag
 +& C^{FP}_{ii-1}\K_{i-1}\left( t_i-t_{i-1}\right)
 \bigg]\,,
\end{align}
where now $t_i$ is the $i^\mathrm{th}$ element on the ordered time grid.
As in the case of the linear grid, the summation can be truncated in order to avoid time regions with low signal. In this case, the summation in Eq.~\eqref{eq:Kselfc-disc-nonl} starts at $j=\max(1,i-i^*)$, where $i^*$ is $i: t^*=t_{i^*}$.

In principle the same procedure may be used with other integration schemes, such as the Simpson scheme, though instabilities may arise depending on how the integration boundaries are treated~\cite{press:07}.
As we show in Appendix~\ref{app:checks}, the Volterra method with a trapezoidal integration scheme is very stable and accurate for small $t$. 

\section{Consistency checks}\label{app:checks}
\begin{figure}[!t]
\includegraphics[width=\columnwidth]{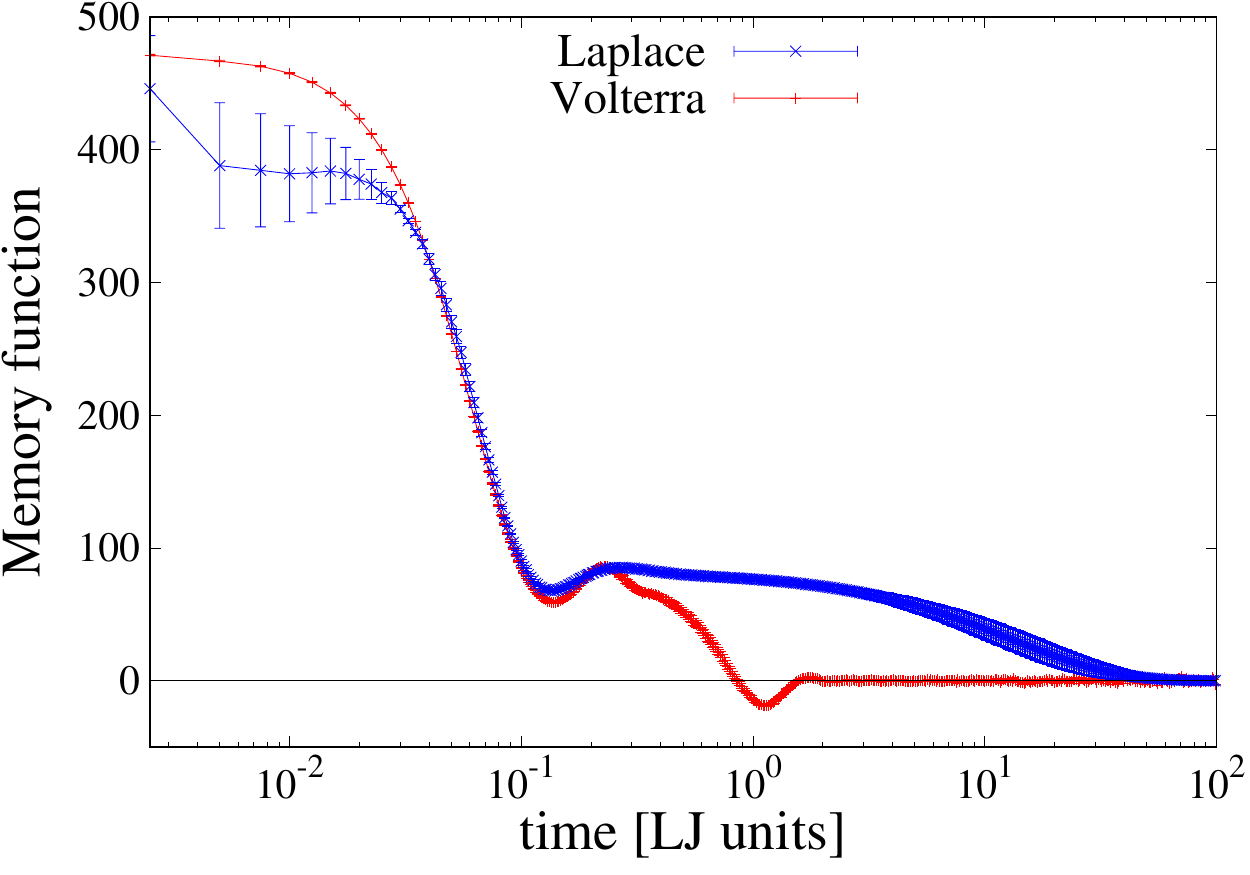}
\caption{Explicit concatenation of the Volterra and Laplace correlation functions, for $T=0.47$.}
\label{fig:concatenate}
\end{figure}
Our method for extracting the memory function consists of the piecewise concatenation of the Volterra method with the Laplace method.
In Fig.~\ref{fig:concatenate} we show this concatenation at $T=0.47$, with jackknife error bars. 
Despite the small fluctuations in the Volterra curves at all temperatures, their behavior becomes unphysical at the end of the short-time regime, and consequently the memory function calculated with the Volterra method fails the test shown at the end of this section.

The Laplace method gives the correct behavior at long times, but suffers from dramatic fluctuations at short times for reasons described in Appendix~\ref{app:laplace}. It is to be noted that at lower temperatures 
the error bars become large with respect to the signal. However, due to systematic sources of error mentioned in Appendix~\ref{app:laplace}, the fluctuations are non-Gaussian, and are thus not a good indicator of the reliability of the calculations.

\begin{figure}[tb]
\includegraphics[width=\columnwidth, trim=55 70 0 0]{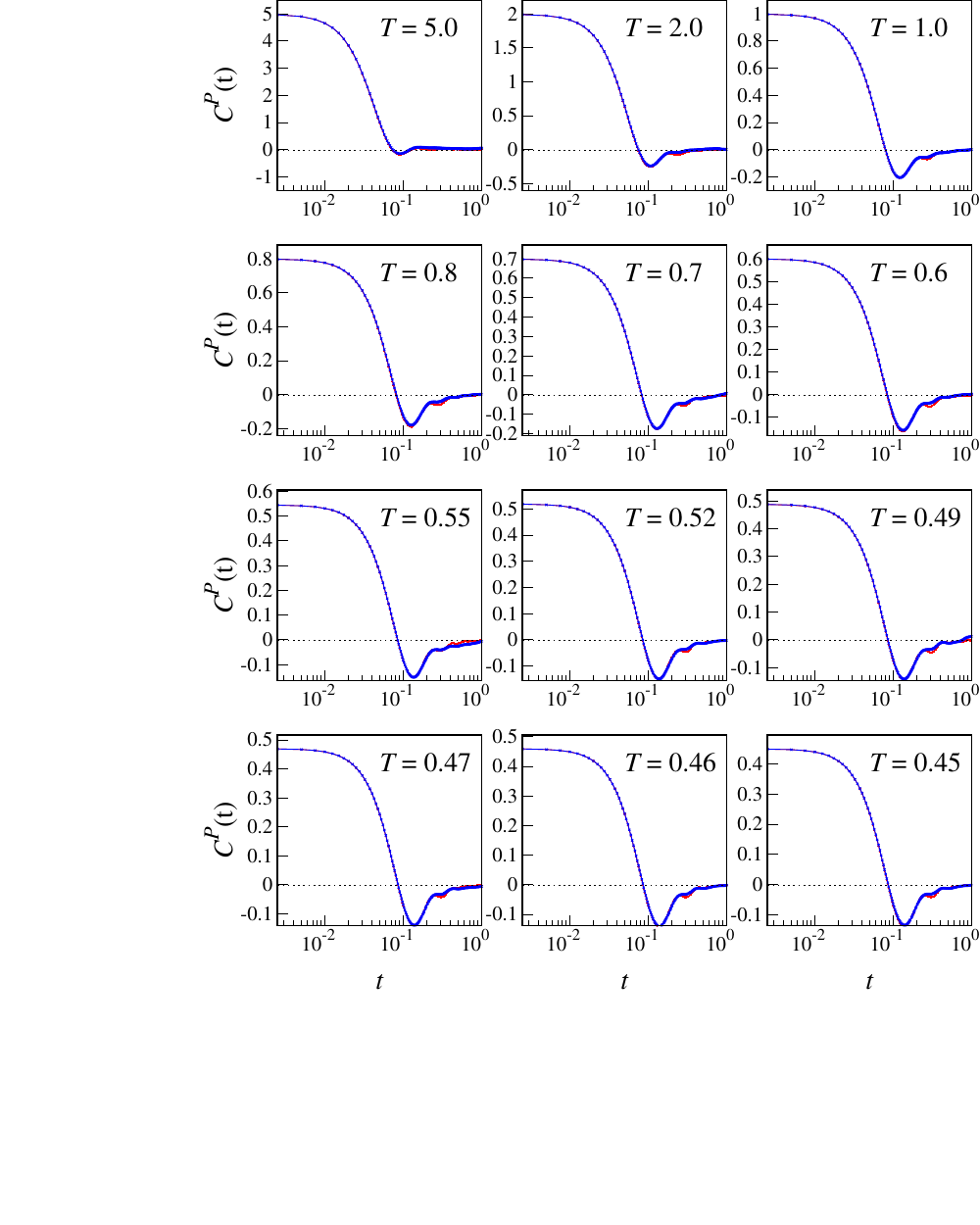}
\caption{Consistency checks for the calculation of the memory function.
In red we plot the measured $\cpp$ as a function of time $t$, and in blue we plot the $\cpp$ calculated from the memory function [Eq.~\eqref{eq:check-cpp}]. The two curves almost coincide.
For each temperature, we plot the left- and the right-hand side of Eq.~\eqref{eq:check-cpp}.}
\label{fig:checks}
\end{figure}
In order to assess the reliability of the computed memory function, we verify that it satisfies Eq.~\eqref{eq:dotcp}.
In Fig.~\ref{fig:checks} we show the result of this consistency check for the integrated version of Eq.~\eqref{eq:dotcp},
\begin{equation}\label{eq:check-cpp}
 \cpp =  - \int_0^t dt \int_0^{t'} \K(t'-u) C^{P}(u) du\,.
\end{equation}
since in our view $\cpp$ can be more readily interpreted by eye than $\cfp$.
We see that the entire non-trivial part of $\cpp$ is well-reproduced. At longer times, the $\cpp$ calculated from the reconstructed memory functions fluctuate wildly, though the time at which the wild fluctuations begin grows as the number of trajectories used for averaging is increased.
clearly, however, for the number of trajectories used here we can achieve full consistency for time scales where $\cpp$ has nearly entirely decayed to zero.

\section{Increasing \texorpdfstring{$\boldsymbol{M}$}{M}}\label{app:M}

\begin{figure}[!tb]
 \includegraphics[width=\columnwidth]{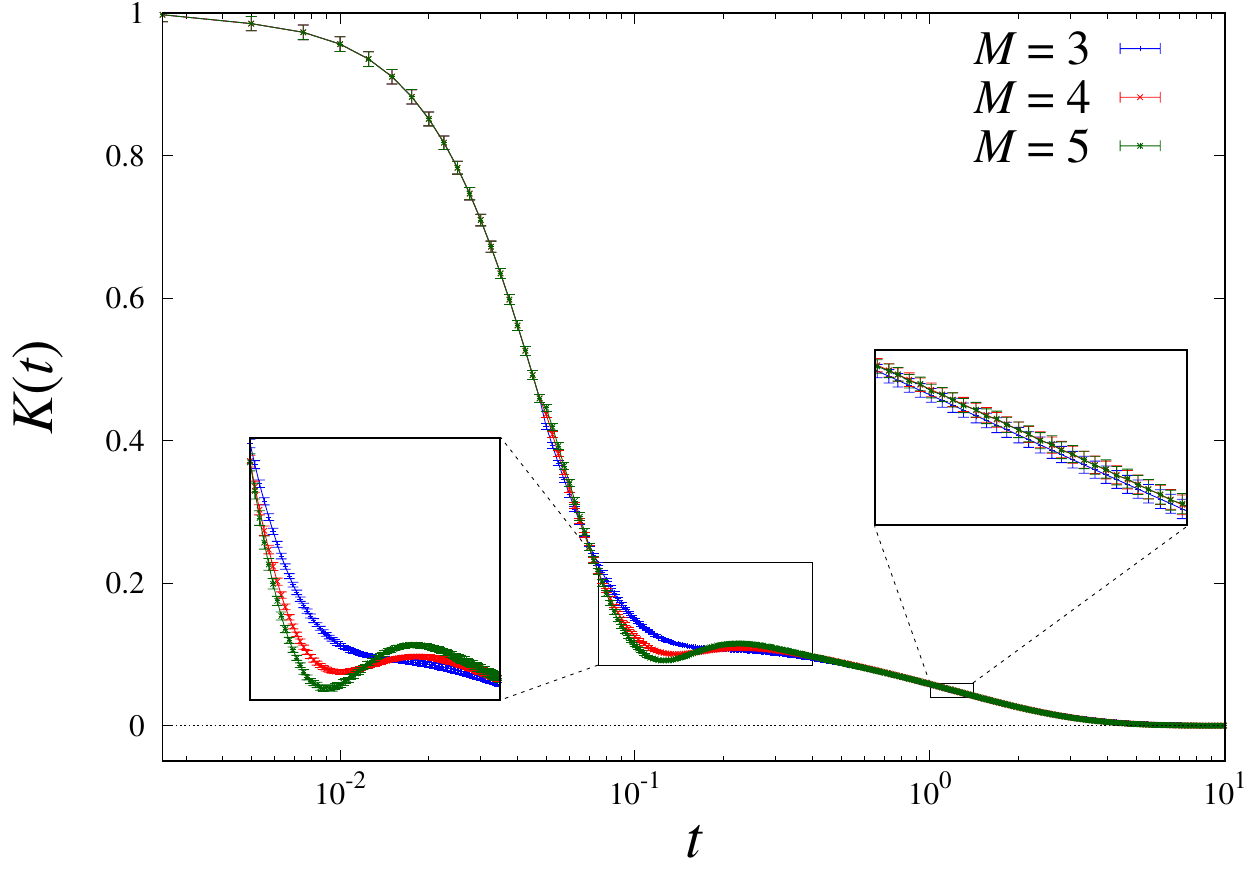}
 \caption{Memory functions at $T=0.8$, calculated for $M=3,4,5$. Zooms of the plateau and of the long-time decay are shown.}
 \label{fig:compareM}
\end{figure}

As stated in Appendix~\ref{app:laplace}, the number of coefficients, $2M$, in the inverse Laplace transform calculated with the Gaver-Stehfest method has an influence on the fine structure of a function that can be resolved. Increasing $M$ allows for the access of finer details in the behavior of $\K(t)$, but doing so also increases numerical instability.

As shown in Fig.~\ref{fig:compareM}, varying $M$ from 3 to 5 produces a variation in the plateau of $\K(t)$, but the long-time behavior remains unchanged. One can expect, therefore, that increasing to $M=7$ (the maximum possible $M$ for double precision, see Appendix~\ref{app:laplace}) will not lead to significant changes. 
Note, moreover, that the plateau heights shown in Fig.~\ref{fig:short}b are evaluated from the long-time part of $\K(t)$ which, as shown in Fig.~\ref{fig:compareM}, does not vary with $M$. Consequently, even though the qualitative shape of the plateau changes, we do not expect our measurements of the plateau height to change if $M$ is increased.

In Fig.~\ref{fig:compareMcheck} we show that the consistency check presented in Appendix~\ref{app:checks} improves as $M$ grows, systematically up to the value of $M$ used in this work.
\begin{figure}[tb]
 \includegraphics[width=\columnwidth]{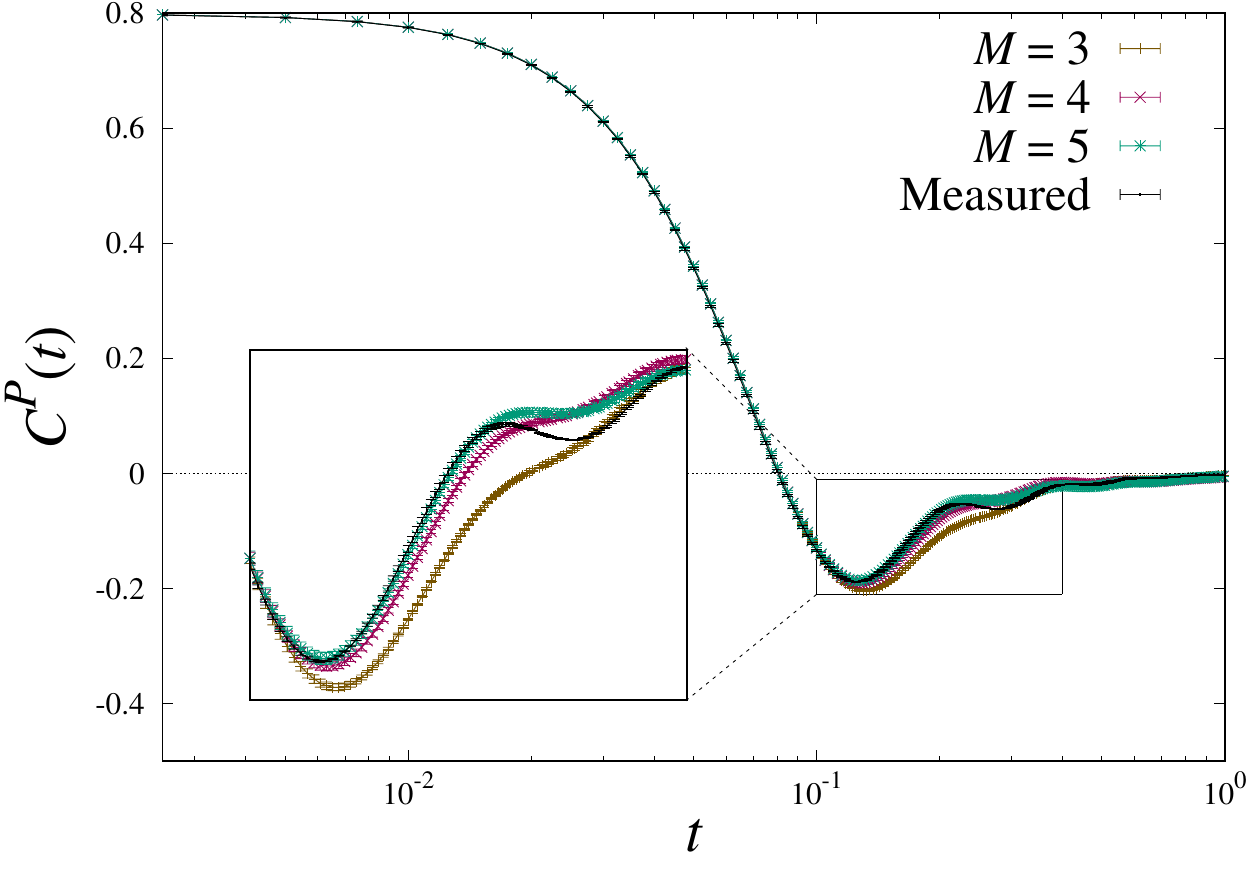}
 \caption{Momentum-momentum autocorrelation function for $T=0.8$, calculated by direct measurement and by integrating the memory functions obtained with $M=3,4,5$.}
 \label{fig:compareMcheck}
\end{figure}

\bibliographystyle{apsrev4-1}
\bibliography{marco}

\end{document}